# Supporting Dynamic Ad hoc Collaboration Capabilities

D. Agarwal and K. Berket
*LBNL, Berkeley, CA 94720, USA*

Modern HENP experiments such as CMS and Atlas involve as many as 2000 collaborators around the world. Collaborations this large will be unable to meet often enough to support working closely together. Many of the tools currently available for collaboration focus on heavy-weight applications such as videoconferencing tools. While these are important, there is a more basic need for tools that support connecting physicists to work together on an ad hoc or continuous basis. Tools that support the day-to-day connectivity and underlying needs of a group of collaborators are important for providing light-weight, non-intrusive, and flexible ways to work collaboratively. Some example tools include messaging, file-sharing, and shared plot viewers. An important component of the environment is a scalable underlying communication framework. In this paper we will describe our current progress on building a dynamic and ad hoc collaboration environment and our vision for its evolution into a HENP collaboration environment.

## 1. INTRODUCTION

Geographically dispersed collaboration teams have been a reality for high energy physicists for many years now. With the approach of turn-on of the Large Hadron Collider at CERN, there has been increased interest in remote access to accelerator facilities and experiments. The CMS and Atlas experiments being built for use at CERN involve collaborations with more than 2000 physicists each and these collaborations are likely to double in size by the time the LHC has beam and data is flowing. For this size collaboration to meet regularly and collaborate in person, would either require all of the physicists to move to a single location or to spend a couple of days each week on airplanes. Internet-based collaboration capabilities are one means of supporting these collaborations and reducing the travel burden while increasing the connection between the collaborators. The various scenarios for usage have included remote operation of an accelerator, remote access to an experiment, and collaborative analysis by a team located at geographically dispersed locations.

Members of the high energy physics community are already adept at providing remote access to meetings using both standard and non-standard videoconferencing technologies. Although this is relatively effective for presentations, it is difficult to provide remote and local participants with an experience of equal participation in interactive discussions. This problem can be caused by several issues including inadequate number and placement of microphones, inability to see all participants, lack of shared applications, inability to have private side conversations, and inability to continue conversations when moving to different rooms (e.g. coffee break or breakout sessions). When videoconferencing is defined as the core component of a collaborative system, working video and audio hardware at every site becomes a requirement. This is impractical and quickly becomes a sociological problem: many people object to being on camera when they don't need to be. Generally it doesn't take long before video cameras are permanently pointed out the window or door rather than at the person.

Although videoconferencing is an important part of supporting formal meetings, engaging in more informal interactions and sharing documents and data have been shown to be an important part of an effective collaboration[17]. Because of this, there is a need for collaborative tools that support connecting people so that they feel like they are working together on a daily, hourly, or even continuous basis[1] [5]. Since high energy physics collaborations are world-wide and often extend beyond normal working hours, support for informal and asynchronous communication capabilities is critical. Collaborators currently revert to email and telephone because present software lacks other asynchronous, informal, and easy to use communication mechanisms. Unfortunately, email messages are often delivered out of sequence and responses from collaborators can become disorganized. Also, the typical scientist now receives so much email during the course of a day that it is ignored during periods of concentrated work and this is often when it is most critical for collaborators to communicate. Messaging systems and presence awareness mechanisms can help to make collaborations successful because they provide basic connectivity in a shared context that allows collaborators to hold spontaneous meetings[14].

Computing activities, including shared program development and debugging and 24x7 data analysis operations, are a central feature of the Atlas and CMS collaboration activities. Grids [10] provide consistent, secure access to distributed computing resources and have gained widespread acceptance as an important future architecture for distributed computing. Grids are becoming the primary environment within which distributed users share computing tasks, analyses, and visualization results. Both Atlas and CMS have committed to using the grid software for their analysis environments. The grid brings a secure environment and opportunities for the development of collaborative tools to interact with computations. It is important that collaborative tools for High Energy Physics integrate with the Grid environment so that users can have a single authentication credential and use single sign on.

## 2. COLLABORATION REALITIES

Existing research and our experience building collaboratories have shown that an effective collaboration framework should provide a consistent environment. The most fundamental characteristic of a collaborative environment is ubiquity. Collaborators should be able to enter and work within the environment from their desktop machine, their laptop, another user's computer, or any other similar device. Users must also be able to participate in collaboration activities from any location (e.g. airport, home, etc.).

Collaboration tools need to be activated in the context of the day-to-day activities of the participants. When collaboration tools are a natural part of working on the computer, a critical mass of collaborators is created in the environment. This makes it possible to find other collaborators easily. If only a subset of the participants login regularly, or if the tools are launched in an out-of-context separate step, collaborators will use them only when they want to contact another user and will generally not find them in the environment. The collaborative tools also need to provide a real benefit to all the users; a one-sided benefit model is counterproductive in a collaboration environment.[11]

Although the extent of the collaboratory tools available to a particular user will depend on the equipment and computing resources available at the user's immediate location, all users must have access to a common core set of collaboratory tools. This core set should provide both the sense and the benefits of connection to the collaboration space. The minimum set of capabilities should include a directory of users, a contact mechanism, and a text-based messaging capability. The minimum user interface should be a web browser. The environment should provide a tangible sense of persistence.

Security is an important characteristic of a collaborative environment. Free and open communication is based on an ability to know the participants' identities. Collaboration tools need an ability to assure participants of the identities of those with whom they are interacting and provide reasonable expectations of privacy during these interactions.

## 3. A BASE TOOLSET FOR HIGH ENERGY PHYSICS COLLABORATORIES

A successful collaborative framework should be adaptable, allowing participants to configure the mode, length and scope of their interactions to meet their own needs. The collaboration environment should work as well for a small group of collaborators as for a large group. Generally, a large collaboration is built from a small group who has convinced others to use the tools. If the tools require a great deal of infrastructure to support them, it is unlikely that the small group would ever have managed to use the tools in the first place. A collaborative framework must support and facilitate a continuum of communication services through an integrated interface so users can easily progress through the continuum and participate in conversations in whatever mode is appropriate at the time. This might mean messaging asynchronously, using an instant messaging system, having an audio conversation, holding a videoconference, sharing an application, or sharing data.

In the remainder of this section we present the components of a basic toolset for HEP collaboratories. In Section 3.1 we take a look at the requirements of the communication substrate for collaboration environments and present protocols that meet these requirements. In Section 3.2 we present a tool that provides a pervasive core contact capability. Finally in Section 3.3 we present a tool for secure sharing of files between collaborators.

### 3.1. A Scalable Communication Substrate

Many current collaboration tools and environments, such as the Access Grid (AG) [19] provide a set of persistent services to users. However, they often rely on a centralized infrastructure. For example, a user wishing to join an AG venue must connect to a server in order to participate in the collaboration. This architecture works well for highly structured collaborations that can afford to run and administer a highly-available server. However, small collaborations are usually built in an ad-hoc manner and often there is no site available to run the server.

At its core, a collaboration environment depends on the collaborators: (1) being able to reliably communicate with each other and (2) knowing the identities of the other collaborators. When the collaboration is conducted over an untrusted network, such as the Internet, security becomes a critical concern. Providing security allows the legitimate collaborators to feel confident about the identities of their partners and securely communicate with them. Although it is possible to establish the communication among the collaborators using unicast mechanisms (e.g. TCP/SSL), this is very complex, inefficient, and difficult to scale. Instead, secure and reliable multicast is a natural underlying communication layer for a collaboration environment.

An instantiation of secure and reliable multicast communication is provided by a combination of the InterGroup protocols [6] [7] and the Secure Group Layer (SGL) [2] [8]. InterGroup is an extension of the TCP concept to the multi-party case that provides membership services, reliable message delivery, and ordered message delivery. The InterGroup protocols are intended to provide these application services in a wide-area environment with a large number of participants, prone to large latencies and frequent faults, such as the Internet. SGL is an extension of the SSL concept to the multi-party case that provides the security services required by applications utilizing reliable multicast communication (e.g. InterGroup) in wide-area environments. SGL establishes secure multicast channels among application components. An SGL secure multicast communication channel is established by first exchanging a session key among the legitimate application components. This key is then used to achieve multicast message confidentiality and/or multicast data integrity within the group. A logical next step is to integrate SGL capabilities into the Grid.

## 3.2. A Pervasive Core Contact Capability

A collaboration environment that is designed to be the core environment for presence, contact, and text messaging is under development at LBNL. It provides a lightweight interface that requires no additional hardware. It is meant to be run continuously on each users desktop. This interface registers the user in the collaboratory and provides availability information and a method of contacting the user. The project developing this software is called the Pervasive Collaborative Computing Environment Project (PCCE).

In order for users to easily locate each other and rendezvous, presence information is provided about people and venues. User information includes information about the user's identity, the user's location, availability, and the venues of which he or she is a member. A view of this information is shown in Figure 1.

Figure 1. PCCE Secure Messaging Client

As part of the PCCE project, a secure messaging application for synchronous and asynchronous messaging is being developed. Within the system users can hold group or one-to-one conversations on an on-going or ad hoc basis. These conversations may be public and open to anyone who is on-line or they may be private and open only by invitation. All conversations take place within venues. To initiate a one-to-one conversation, users create a private venue to which they invite someone else. The conversation can be extended to a larger group by its members making the venue public or explicitly inviting others. Users may participate in multiple discussions simultaneously and they may leave notes for others who are on-line or off-line.

The current implementation is based on a client-server model that supports client and server authentication and encryption of messages exchanged over the network. In order to leverage existing technologies, we modified a public domain IRC server (IRCD hybrid) to replace its nonsecure TCP sockets with SSL connections. To provide persistence (e.g., unique nicknames and permanent venues) and enhanced presence information independent of any one chat environment, we developed a custom PCCE server which also provides authentication and authorization services. Both the IRC and PCCE servers use only SSL network connections and have their own X.509 credentials which are presented to each other and to clients.

To access the servers, users must pre-register with the PCCE server through either a designated system administrator or a registered user with administrative privileges. After having registered, users log into the system via username and password and can then use the client interface to edit their own personal information and register their distinguished names in their X.509 certificates. Subsequent login can be by either certificate or username and password, and users who authenticate by certificate are granted extended privileges (e.g, the ability to create new user accounts and permanent venues). The registration process facilitates asynchronous messaging as well as access control. The PCCE server stores registration information in a local database and uses the information to make authorization decisions (e.g., who can connect to the IRC server, who can leave and receive notes, and who can perform administrative operations).

This design forces users of PCCE to rely on this server. PCCE is extending its security architecture based on an incremental trust model [4] in order to allow users to spontaneously create a secure collaboration, join an on-going collaboration or single session without prior registration. With these changes and a migration to using an InterGroup and SGL core for communication so the collaboration will be able to operate without either server. By removing the dependence on the servers, PCCE will gain the ability to run in a purely ad-hoc manner. The PCCE server, if present, will enhance the functionality of the collaboration, rather than creating a (security) bottleneck, and the IRC server can be removed completely.

## 3.3. File-sharing

Groups collaborating on scientific experiments have a need to share information and data. This information and data is often represented in the form of files and database entries. In a typical high energy physics collaboration, there are many different locations where data would naturally be stored. This makes it difficult for collaborators to find and access the information they need. The goal of *scishare* [30] is to create a lightweight information-sharing system that makes it easy for collaborators to find and use the data they need. This system must be easy-to-use, easy-to-administer, and secure.

The *scishare* information sharing tool is in many ways similar to existing peer-to-peer file sharing systems, such as Gnutella [15], Kazaa [28], Limewire [29], etc. Each peer designates a set of items to share within the system.

Peers are able to search for items by sending a query to the network. The network delivers this query to the other peers, which run the query against the items they have designated to share. Metadata about the matching items is sent back to the peer that originated the query. This metadata contains information about the items, as well as how the actual items may be retrieved. Then the peer may go ahead and retrieve the item.

As part of our design, we allow for easy configuration of the message exchange patterns of the system. Currently, we use the InterGroup protocols to reliably deliver each query to all of the current participants in a scalable manner, without having to discover all of their identities. The transfer of the metadata and items is done using HTTP. Just as easily, we could transfer the metadata using the InterGroup protocols. We plan to make use of this configurability to perform measurements on the performance and resource consumption of different message exchange patterns.

In our design we also separated the application-specific requirements of information sharing from the more generic discovery mechanisms. We hope to advance and reuse the discovery component in other projects.

The prototype implementation of *scishare* is a Java application with a graphical front-end. It provides the user the ability to search for remote files and view searches being run on their machine (Figure 2), to transfer remote files to the local machine, to manage the locally shared files (Figure 3), and to monitor the transfer of files to and from their machine (Figure 4).

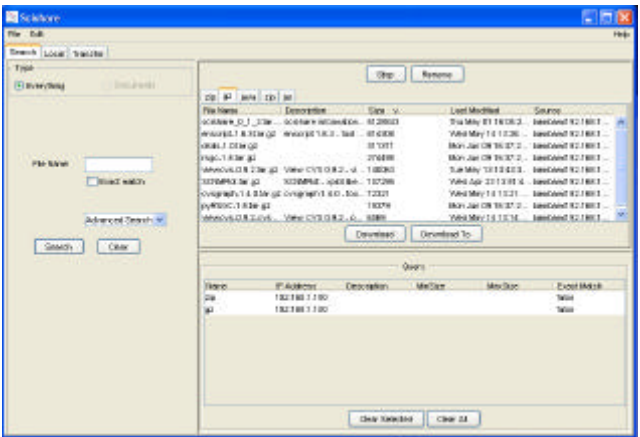

Figure 2. Performing a search in the *scishare* application.

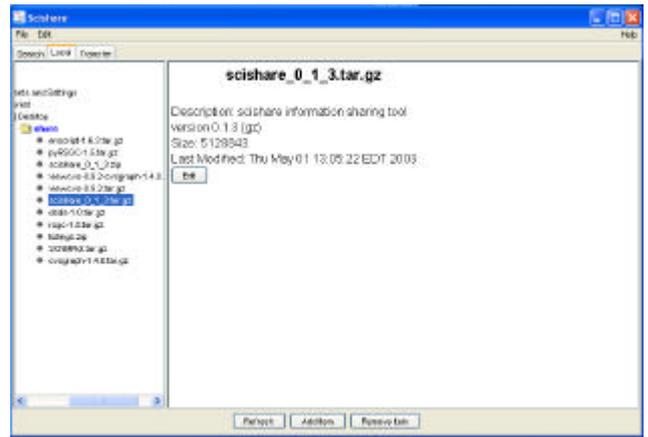

Figure 3. Managing the locally shared files in the *scishare* application.

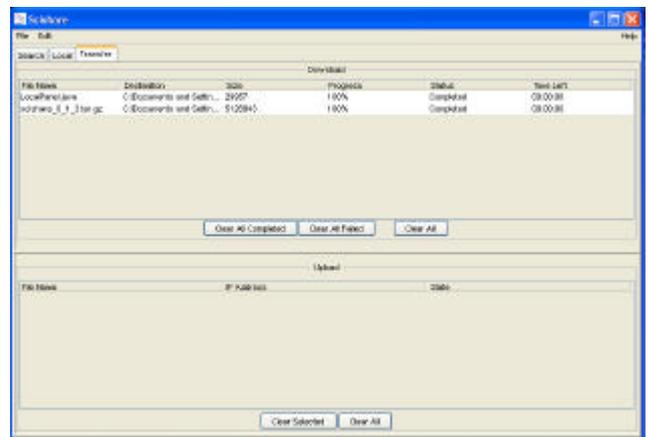

Figure 4. Monitoring file transfers in the *scishare* application.

We are currently in the process of incrementally adding security to *scishare*. We first use the concept of interposing a security layer between the application and the transport layer protocol in order to secure the communication. This means we will be using HTTPS instead of HTTP an SGL instead of InterGroup. Our design allows us to do this with little impact on the existing application code. Once complete, this will provide confidentiality, integrity, authenticity, and implicit authorization enforcement for the peer communication.

The next step is to tackle the issue of access control to the other resources, e.g. files. Our system will allow for individual peers to use the authorization services of their choice. The implementation that we provide will make use of Akenti distributed authorization [16] for this service. Akenti targets widely distributed environments where resource owners and users could span many autonomous organizations. It provides a powerful authorization policy language the enables fine-grained access control to resources in dynamic collaborations.

## 4. RELATED WORK

Many tools have been developed for online collaboration and in this section we review some of the tools relevant to scientific collaboration. These include tools and systems for messaging, on-line instrument access, and videoconferencing. These tools tend to target specific interaction modes rather than a continuum of interaction paradigms.

Among the text-based messaging systems that have been developed over the past several years is Internet Relay Chat (IRC) [20]. The IRC metaphor for rooms is channels; anyone can create a channel and it only exists while someone is in the channel. Recent systems such as the America On-Line Instant Messaging program (AIM) [22] and the ICQ messaging system[23] are primarily intended for one-to-one conversation. Later versions of AIM include group chat, file transmit, and voice capabilities. Businesses have begun to use messaging tools for informal communication in the workplace. Recent research efforts[14] have found that the use of these tools contributes to a sense of co-worker community often lacking in a large distributed workplace. The tools are also seen as alleviating some of the isolation and fragmentation that accompanies telecommuting or constant travel. Lotus and Microsoft have developed their own versions of the instant message service and the Lotus system supports secure private messaging between authenticated users.

Several systems such as the Virtual Room Videoconferencing System (VRVS) [21] and commercial H.323 systems offer videoconferencing capabilities. These tools provide conventional two-way videoconferencing and broadcast of one primary video and audio feed to an audience. The MBone videoconferencing tools [24] (vic, vat, rat, wb, and sdr) provide multicast-based, multi-way videoconferencing that allows all users to be seen and heard as equal participants in the videoconference. The Access Grid [19] has expanded on this idea by providing relatively natural group-to-group interaction capabilities and launching mechanisms that allow many groups to participate simultaneously. Recently the VRVS system has been enhanced to include interaction with MBone and Access Grid videoconferencing sessions.

Peer-to-peer technologies for distributed file management either deal with distributed storage or file-sharing. OceanStore [13] and Chord [9] provide highly available distributed storage through the use of distributed hashtables. Gnutella [15], Kazaa [28], Limewire [29] and similar systems allow for sharing of files between a large and dynamic set of participants.

Groove [25] facilitates group work by supporting instant messaging, file-sharing, web-browsing, and forum discussions. As a composable framework, it offers a set of customizable tools. Based on a peer-to-peer paradigm, there is no need for a centralized server. Groove also supports offline use, message encryption, and persistence across computers.

The US Department of Energy has sponsored several collaboratory tool research and development projects including development of an electronic notebook [26], a collaboratory interoperability framework [3], the CORE2000 system, the Akenti authorization server [16], and Access Grid components. Concurrent with development of these underlying technologies, DOE has also sponsored the development of several pilot collaboratories in specific scientific disciplines including a virtual NMR facility [12], the Materials Microcharacterization Collaboratory [18], the Diesel Combustion Collaboratory [27], and the SpectroMicroscopy Collaboratory [5]. These early projects provided important insights into the scientific collaboration environment.

## 5. CONCLUSION

High energy physics collaboratories will benefit from the base toolset we have presented. By using InterGroup and SGL as core communication services in the collaboration environment, existing collaborations can easily operate in either an ad hoc or infrastructure-enabled setting without sacrificing security. The pervasive core contact capability allows the collaborators to easily locate each other and communicate in a synchronous or asynchronous manner. The file sharing capability allows the collaborators to exchange data in a convenient, easy to use and secure manner.

This base toolset allows collaborations to function without the dependence on any single server. This allows servers to provide added value services rather than being essential components. Thus, the dependence on centralized infrastructure is reduced and informal, spontaneous collaborations are enabled.

### Acknowledgments

This work was supported by the Director, Office of Science, Office of Advanced Computing Research, Mathematical Information and Computing Sciences Division, of the U.S. Department of Energy under Contract No. DE-AC03-76SF00098. This document is report LBNL Report number LBNL-53355.

### References

[1] D. Agarwal, ``Collaborating Across the Miles", Proceedings INMM/ESARDA Workshop on Science and Modern Technology for Safeguards, Albuquerque, NM, Sept. 1998.

[2] D.A. Agarwal, O. Chevassut, M.R. Thompson and G. Tsudik, "An Integrated Solution for Secure Group Communication in Wide-Area Networks," Proceedings IEEE Symposium on Computers and Communications, July 2001.

[3] D. Agarwal, I. Foster and T. Strayer, "Standards-Based Software Infrastructure for Collaborative Environment and Distributed Computing

Applications," white paper available at http://www-itg.lbl.gov/CIF/.

[4] D. Agarwal, M. Lorch, M. Thompson, M. Perry, "A New Security Model for Collaborative Environments", Proceedings of the Workshop on Advanced Collaborative Environments, Seattle, WA, June 2003.

[5] D. A. Agarwal, S. R. Sachs, and W. E. Johnston, "The Reality of Collaboratories," Computer Physics Communications vol. 110, issue 1-3 (May 1998).

[6] K.Berket, D.A. Agarwal, O. Chevassut, "A Practical Approach to the InterGroup Protocols", Future Generation Computer Systems, Vol. 18 (5), Elsevier Science B.V., 2002, pp. 709-719.

[7] K. Berket, D. A. Agarwal, P. M. Melliar-Smith and L. E. Moser. "Overview of the InterGroup Protocols." Proceedings International Conference on Computational Science, San Francisco, CA.

[8] E. Bresson, O. Chevassut and D. Pointcheval, *"The Group Diffie-Hellman Problems,"* Proceedings of Selected Areas in Cryptography (SAC'02), St John's, Newfoundland, Canada, August 15 - 16, 2002.

[9] F. Dabek, E. Brunskill, F. Kaashoek, D. Karger, R. Morris, I. Stoica, and H. Balakrishnan, " Building Peer-to-Peer Systems With Chord, a Distributed Lookup Service," Proceedings of the 8th Workshop on Hot Topics in Operating Systems (HotOS-VIII), Schloss Elmau, Germany, May 2001.

[10] I. Foster and C. Kesselman (eds.), The Grid: Blueprint for a New Computing Infrastructure, Morgan Kaufmann, 1999.

[11] J. Grudin, "Groupware and Social Dynamics: Eight Challenges for Developers." Communications of the ACM, Vol. 34(1), Jan. 1994.

[12] K. Keating, J. Myers, J. Pelton, R. Bair, D. Wemmer, and P. Ellis, "Development and Use of a Virtual NMR Facility," Journal of Magnetic Resonance, December 1999.

[13] J. Kubiatowicz, D. Bindel, Y. Chen, S. Czerwinski, P. Eaton, D. Geels, R. Gummadi, R. Rhea, H. Weatherspoon, W. Weimer, C. Wells, and B. Zhao, "OceanStore: An Architecture for Global-Scale Persistent Storage," Proceedings of the Ninth international Conference on Architectural Support for Programming Languages and Operating Systems (ASPLOS 2000), November 2000.

[14] B. Nardi, S. Whittaker, E. Bradner, "Interaction and Outeraction: Instant Messaging in Action," Proceedings Conference on Computer Supported Cooperative Work, Philadelphia, PA (Dec. 2000).

[15] Peer-to-Peer: Harnessing the Power of Disruptive Technologies, A. Oram, editor, O'Reilly & Associates, Inc., Sebastopol, CA, 2001.

[16] M. Thompson, A. Essiari, S. Mudumbai, "Certificate-based Authorization Policy in a PKI Environment", to appear in ACM Transactions on Information and System Security.

[17] S. Whittaker, D. Frohlich, and W. Daly-Jones, "Informal Workplace Communication: What is it Like and How Might We Support It?" Proceedings of CHI'94 Conference on Human Factors in Computing Systems.

[18] N. Zaluzec, "Interactive Collaboratories for Microscopy & Microanalysis," Journal of Computer Assisted Microscopy, Oct. 1997.

[19] http://www-fp.mcs.anl.gov/fl/accessgrid/.
[20] http://www.irc.org/
[21] http://vrvs.cern.ch/ or http://www.vrvs.org/
[22] http://www.aim.com/
[23] http://web.icq.com/
[24] http://www-itg.lbl.gov/mbone/
[25] http://www.groove.net/
[26] http://www.csm.ornl.gov/enote/
[27] http://www-collab.ca.sandia.gov/
[28] http://www.kazaa.com
[29] http://www.limewire.com
[30] http://www-itg.lbl.gov/P2P/file-share/